# Understanding and Mitigating Harmful Design in User-Generated Virtual Worlds[*]

Zinan Zhang, Xinning Gui, & Yubo Kou

College of IST, Penn State, USA, zzinan@psu.edu, xinninggui@psu.edu, yubokou@psu.edu

Virtual space offers innovative ways for individuals to engage with one another in a digital setting. Prominent virtual social platforms, such as Facebook Spaces, VR Chat, and AltspaceVR, facilitate social connections, allowing users to interact seamlessly. Additionally, certain video games, like Second Life and World of Warcraft, are set within these virtual spaces as well, providing immersive player experiences. As the popularity of virtual space grows, various companies have begun to democratize the process of creating these spaces, shifting the development from skilled professionals to hobbyist creators. Platforms like Minecraft, Roblox, and RecRoom enable users to create and publish their own virtual environments, hosting a wide range of interactions and narratives. This shift echoes the rise of user-generated content, where content creators create and publish content on platforms, such as social media platforms [6]. For example, YouTubers upload videos on YouTube and Reddit users post text-based content on Reddit. For a long time, user-generated content has predominantly contained text, videos, and images. However, with the emergence of virtual spaces, some platforms now allow creators to create and publish their own virtual spaces, leading to the emergence of user-generated virtual worlds.

User-generated virtual worlds (UGVWs) offer a multitude of experiences within these virtual spaces. For example, Roblox, managed by Roblox Corporation, has over 40 million UGVW (termed 'experiences' by Roblox) created by creators, which attracts over 214 million monthly players, 85% of whom are younger than 18. The virtual spaces on Roblox vary from live concerts to a variety of gameplay genres. Roblox has hosted several virtual concerts, attracting attendance of up to 33 million times. Games like National Hockey League (NHL) Blast offer players engaging experiences in playing hockey with others, accumulating more than 18 million visits and over 1.4 million hours of gameplay. Yet, not all interactions or experiences within these virtual spaces are beneficial or playful, and some can be harmful. For example, Roblox has faced criticism for hosting content with problematic themes such as terrorism and virtual sexual assault (see more examples in [1]) and grabbing cash from children [5]. As such, UGVW such as Roblox may present significant risk and harm, especially to its predominantly young audiences, including but not limited to propagating terrorist ideologies, causing trauma through sexually explicit interactions, and resulting in financial loss due to the extensive spending by children.

To explore emergent forms of harm present in UGVW, our research team has conducted preliminary work on Roblox [1]. Utilizing a grounded theory approach, we were able to generate a taxonomy of harmful design patterns as experienced by players in UGVWs on Roblox. Distinct from harm in traditional forms of user-generated content, harmful designs in UGVW introduce novel forms of harm. In contrast to video, text, or image content, where harm can be directly captured, interpreted, and perceived by the victim, we found that UGVWs can subtly manipulate users into certain behaviors, such as making purchases through prompts. In addition, these virtual worlds may feature unmoderated social designs that inappropriately engage users in specific actions within the virtual space. Users are also exposed to dynamic and emergent risks through exposure to unregulated content in real time. Lastly, problematic UGVW designs, such as those in storytelling, can encourage users' behaviors in virtual spaces aligned with extremist ideologies, values, and ideas.

---

[*] We thank NSF No. 2326505 for supporting the work.

There could be many factors that cause the emergence of harm in UGVWs. For example, users in the UGVW could facilitate harm through interactions with other users in the same world as them, such as those who conduct 'sexual assault.' Creators created the design of social interactions and content within these games, providing a space and ability for harmful content and harmful interactions. Roblox Corporation is criticized for exploiting child labor [2] due to its revenue-sharing model with creators, which might be the reason for integrating harmful UGVW designs, like ubiquitous microtransactions, into their virtual worlds to secure enough income. In addition, the resources provided to creators in the Roblox community such as models may contain viruses and lead to harmful designs in the virtual worlds. Moderation and detection of harm are currently not enough for capturing all types of harm in different ways such as in the storytelling, models, or interactions. Therefore, we believe addressing harm in UGVWs requires multiple focuses rather than solely focusing on one aspect.

To effectively mitigate harm in UGVWs, it is crucial to implement appropriate moderation strategies that assist creators in developing less harmful designs and users in conducting inappropriate behaviors. Findings from our recent research [4] indicate that moderation alone is insufficient in curbing toxic behaviors in video games. A rehabilitative approach, offering support for players to adopt fewer toxic behaviors with community backing, may be more effective. Similarly, merely penalizing creators or users may not deter the emergence of harm in UGVW designs. We envision that future research should explore methods to rehabilitate and educate creators, guiding them toward creating positive and constructive virtual world designs.

Our current and ongoing work aligns with multiple topics of this workshop. We look forward to having productive conversations with like-minded researchers to better understand new forms of harm in emerging virtual spaces and identify effective strategies to mitigate those harms.